\documentclass[runningheads]{llncs}
\usepackage[utf8]{inputenc}
\usepackage[dvipsnames]{xcolor}
\usepackage{graphicx}
\usepackage{multirow}
\usepackage{enumitem}
\usepackage{verbatim}
\usepackage[para]{footmisc}
\usepackage{tikz}
\usepackage{pgfplots, pgfplotstable}
\pgfplotsset{compat=1.7}
\usepackage{subcaption}
\usepackage{hhline}
\usepackage[hyphens]{url}

\usepackage{array}
\DeclareRobustCommand\legendbox[1]{(\textcolor{#1}{#1}~\begin{tikzpicture}[x=0.2cm, y=0.2cm] \draw [color=black, fill=#1!20] (0,0) -- (0,1) -- (0.6,1) -- (0.6,0) -- (0, 0); \end{tikzpicture})}
\DeclareRobustCommand\LegendBox[2]{(\textcolor{#1}{#2}~\begin{tikzpicture}[x=0.2cm, y=0.2cm] \draw [color=black, fill=#1!40] (0,0) -- (0,1) -- (0.6,1) -- (0.6,0) -- (0, 0); \end{tikzpicture})}

\pgfplotsset{
SmallBarPlot/.style={
    font=\footnotesize,
    ybar,
    width=\linewidth,
    ymin=0,
    xtick=data,
    xticklabel style={text width=0.8cm, align=center},
    xtick pos=left,
},
BlueBars/.style={
    fill=MidnightBlue!70, bar width=0.25
},
RedBars/.style={
    fill=red!40, bar width=0.25
}
}

\usepackage[normalem]{ulem}

\usepackage[disable]{todonotes}

\newcommand{\TODO}[1]{\todo[color=orange!10,linecolor=black!50]{\textbf{TODO:} #1}}
\newcommand{\Aron}[1]{\todo[color=yellow!5,linecolor=black!50]{\textbf{Aron}: #1}}
\newcommand{\Ayman}[1]{\todo[color=blue!5,linecolor=black!50]{\textbf{Ayman}: #1}}
\newcommand{\Shanto}[1]{\todo[color=green!5,linecolor=black!50]{\textbf{Shanto}: #1}}
\newcommand{\Amin}[1]{\todo[color=red!5,linecolor=black!50]{\textbf{Amin}: #1}}

\newif\ifSeparatePages
\SeparatePagesfalse
\newcommand{\ClearPage}{\ifSeparatePages\clearpage\else\fi}

\newcommand{\SPACE}[1]{}

\newcommand{\extended}[1]{#1}
\newcommand{\notExtended}[1]{}

\title{Smart Contract Development from the Perspective of Developers: 
Topics and Issues Discussed on Social Media}
\titlerunning{Smart Contract Development from the Perspective of Developers}

\author{
Afiya Ayman \and
 Shanto Roy \and
 Amin Alipour \and
 Aron Laszka
}
\institute{University of Houston}

\begin{document}


\setlength{\marginparwidth}{4.3cm}

\maketitle

\begin{center}
Accepted for publication in the proceedings of the\\4th Workshop on Trusted Smart Contracts (WTSC) in association with Financial Cryptography (FC 2020).
\end{center}

\begin{abstract}
Blockchain-based platforms are emerging as a transformative technology that can provide reliability, integrity, and auditability without trusted entities.
One of the key features of these platforms is the trustworthy decentralized execution of general-purpose computation in the form of smart contracts, which are envisioned to have a wide range of applications.
As a result, a rapidly growing and active community of smart-contract developers has emerged in recent years.
A number of research efforts have investigated the technological challenges that these developers face, introducing a variety of tools, languages, and frameworks for smart-contract development, focusing on security.
However, relatively little is known about the community itself, about the developers, and about the issues that they face and discuss.
To address this gap, we study smart-contract developers and their discussions on two social media sites, Stack Exchange and Medium.
We provide insight into the trends and key topics of these discussions, into the developers' interest in various security issues and security tools, and into the developers' technological background.
\end{abstract}

\thispagestyle{plain}
\pagestyle{plain}

\section{Introduction}
\label{sec:intro}

\newcommand{\draft}[1]{\textcolor{blue}{#1}}

The popularity and adoption of blockchain based platforms are growing rapidly both in academia and industry.
This growth is driven by the unique features of blockchains: providing integrity and auditability for transactions in open, decentralized systems.
While earlier blockchains, such as Bitcoin~\cite{nakamoto2008bitcoin}, used these features to establish cryptocurrencies, more recent blockchains, such as Ethereum, also function as distributed computational platforms~\cite{underwood2016blockchain,wood2014ethereum}. 
These platforms enable developers to deploy general-purpose computational code in the form of smart contracts, which can then be executed by a decentralized but trustworthy system. 
Smart contracts are envisioned to have a range of innovative applications, such as asset tracking in the Internet of Things~\cite{christidis2016blockchains}, privacy-preserving transactive energy systems~\cite{laszka2018transax,wang2019when}, and various financial applications~\cite{tapscott2017blockchain}.

Since the security of these applications hinges on the correctness of the underlying contracts, it is crucial that developers are able to create correct contracts.
Sadly, the development of smart contracts has proven to be a challenging and error-prone process, in large part due to the unusual semantics of smart contract platforms and languages~\cite{atzei2017survey,luu2016making}.
Studies have found that a large number of contracts that are deployed on the main Ethereum network suffer from various security issues~\cite{luu2016making,nikolic2018finding}.
Such issues may manifest as security vulnerabilities, some of which have led to security incidents with financial losses in the range of hundreds of millions of dollars worth of cryptocurrencies~\cite{finley2016million,newman2017security}.
%
As a response, the research community has stepped forward and introduced a number of tools (e.g., \cite{angelo2019survey,luu2016making,tsankov2018securify,nikolic2018finding}), frameworks (e.g., \notExtended{\cite{mavridou2018designing,mavridou2019verisolid}}\extended{\cite{mavridou2018designing,mavridou2019verisolid,mavridou2018tool}}), and even new languages (e.g., \cite{oconnor2017simplicity}) to help developers.

While the technical capabilities of these tools and frameworks have been evaluated by multiple surveys (e.g., \notExtended{\cite{di2019survey,kirillov2019evaluation}}\extended{\cite{di2019survey,kirillov2019evaluation,chen2019survey,liu2019survey,harz2018towards}}), relatively little is known about whether developers use them in practice or even whether developers are aware of them. 
In fact, to the best of our knowledge, no prior work has studied the smart contract developers' awareness of security issues and tools or about which issues they are most concerned.
%
In light of this, there is a clear gap in research regarding the developers' perspective of smart contract development. 
Further, very little is known about the developers' technological background and interests, and about their online communities. 
Such information is crucial for enabling researchers to better understand the potential entry barriers for smart contract technology and for guiding researchers to address the developers'~needs.

To address this gap, we study the smart contract developers' online communities, the  topics that they discuss, and their interest in various security issues and tools.
To this end, we collect data from three social media sites: \emph{Stack Overflow}, the most popular Q\&A site for software developers \notExtended{\cite{Barua2014}}\extended{\cite{Barua2014,BestActiveForumsProgramming,Top10sitesProgramming}}; \emph{Ethereum Stack Exchange}, a site focusing on Ethereum from the leading network of Q\&A sites~\cite{Ethereum82:online}, and \emph{Medium}, a popular blog hosting site~\cite{HowMediu89:online}.
In particular, we collect and analyze discussions about smart contracts (e.g., posted questions, answers, blog entries, comments) as well as information about the users who participate in these discussions. We seek to answer the following research questions:
%

\begin{itemize}[leftmargin=2em, topsep=0.0em]
    \item[\textbf{Q1}] \textbf{Trends:} What are the main trends in smart contract related discussions? How do they compare to discussions related to other technologies?
    \item[\textbf{Q2}] \textbf{Security:} Which common security issues and tools do developers discuss? Do discussions about security issues and tools coincide? 
    \item[\textbf{Q3}] \textbf{Developers:} What are the smart contract developers' technological background and interests besides smart contracts? 
    
\end{itemize}

We answer the above questions in our analysis (Section~\ref{sec:results}); here, we highlight a few interesting results.
We find that the intensity of smart contract related discussions reached its peak in early 2018 and has been slowly declining since then (while discussions about other technologies have remained stable). This coincides with the decline of ETH price, which peaked in January 2018. 
In the terminology of the so-called `hype cycle'~\cite{fenn2008mastering},
this suggests that
smart contracts may have passed the `peak of inflated expectations' and are now in the `trough of disillusionment' phase.
This is in interesting contrast with a 2019 July Gartner report~\cite{litan2019hype},
which placed smart contracts at the peak of expectations.
On Stack Overflow and Ethereum Stack Exchange, we find that most questions about smart contracts receive at least one answer, while the majority of questions about other technologies remain unanswered; however, questions about smart contracts are less likely to lead to lengthy discussions.
We also find that very few discussions are related to security, with re-entrancy being the most discussed vulnerability (in part due to the so-called ``DAO attack'', as evidenced by our findings).
There are even fewer mentions of security tools, even in security related discussions.
However, we find a significantly higher number of security related posts on Medium.
On all sites, smart contract related discussions are dominated by a few key technologies and languages (e.g., Solidity, web3.js, Truffle).
Besides smart contracts, the topics that are most discussed by smart contract developers on Stack Overflow are related to web (e.g., jQuery, HTML, CSS).
On Medium, we find that smart contract developers are also interested in non-technical topics related to entrepreneurship (e.g., startups, finance, investing).

\paragraph{Outline}
\TODO{Check and update before submission!}
The remainder of this paper is organized as follows.
In Section~\ref{sec:study}, we describe our data collection and analysis methodology.
In Section~\ref{sec:results}, we present the results of our study.
In Section~\ref{sec:related}, we give a brief overview of related work.
Finally, in Section~\ref{sec:concl}, we discuss our findings and provide concluding remarks.

\ClearPage
\section{Study Design}
\label{sec:study}

\subsection{Data Collection}
\label{sec:data}

\subsubsection{Research Ethics}  
Our study is based on publicly available data, and we report statistical results that contain no personally identifiable information.

\subsubsection{Stack Exchange}
\Aron{It would be good to provide stronger motivation for using these data sources}
is a network of question-and-answer (Q\&A) websites.
%
We collect data from two Stack Exchange sites:
\emph{Stack Overflow\footnote{\url{https://stackoverflow.com/}}}, the most popular generic site for developers~\notExtended{\cite{Barua2014}}\extended{\cite{Barua2014,BestActiveForumsProgramming,Top10sitesProgramming}},
and
\emph{Ethereum Stack Exchange\footnote{\url{https://ethereum.stackexchange.com/}}}, the site that focuses on Ethereum.
%
On these two websites, posts have the same structure:
each post includes a question, a title, a set of associated tags, a set of answers, and a set of comments. 
Only registered users can post new questions or answer existing ones, which enables us to study the developers. 
\Aron{Perhaps include some parts of this here:
``In all Stack Exchange website, the registered users have a unique AcountId which is identical among all sites.  Ethereum Stack Exchange was launched in 2016, whereas Stack Overflow started in 2008. Many of the users of the Ethereum Stack Exchange are also users of Stack Overflow. 
\SPACE{As of September 20th, 2019, Stack Exchange has 11,013,825 registered  users~\cite{StackExchangeDataExplore} for both Ethereum Stack Exchange and Stack Overflow.} 
We see that compared to other users, a larger fraction (48\%) of the smart contract developers' profiles are  new (i.e., created after the year 2017). ''}
To facilitate searching and categorizing posts,
Stack Exchange requires users to associate one or more tags with each question. 
These tags are unstructured and chosen by the users, so they include a wide range of terms (e.g., \emph{Python}, \emph{linked-list}).

From Ethereum Stack Exchange, we collect all posts and users using Stack Exchange Data Explorer (SEDE)~\cite{StackExchangeDataExplore}.
We also collect all posts and users from Stack Overflow using the quarterly archives hosted on the Internet Archive~\cite{stackexchangeInternetArchive}, which we complement with the latest data from SEDE. 
Since Stack Overflow is a generic site for developers, we need to find 
posts that are related to smart contracts.
To this end, we use a snowballing methodology. 
First, we find all posts whose tags contain \emph{smartcontract}. 
Then, we extract other tags from the collected posts, identify the most frequently used tags that are strictly related to smart contracts, and extend our search with these tags.
We continue repeating this process until we cannot collect any more related posts.
In the end, we search for posts whose tags contain the following strings (except for \emph{ether}, which needs to be an exact match to avoid finding, e.g., \emph{ethernet}):
\emph{smartcontract}, 
\emph{solidity},
\emph{ether}, 
\emph{ethereum}, 
\emph{truffle}, 
\emph{web3}, 
\emph{etherscan}.
Finally, we manually check a random sample of the collected posts to confirm that they are indeed related to smart contracts.
In total, we collect 30,761 smart contract related questions, 38,152 answers, and 73,608 comments as well as the 56,456 users who posted these.
Our dataset includes everything up to November 22, 2019.



\subsubsection{Medium}\hspace{-0.6em}\footnote{\url{https://medium.com/}}
is a popular blog platform  \cite{HowMediu89:online}, where registered users can publish posts on a variety of subjects, and other users may read, respond (i.e., comment), clap, or vote.
%
A Medium post \Aron{Not always?}typically contains a title, a text body, tags, reader responses, number of claps and votes, author's name and profile URL, reading time based on word count, and publication date\Aron{We should probably list only the things that we use in our study}. 
Since Medium is a generic blog site, we again use a snowballing methodology to collect smart contract related posts, similar to Stack Overflow.
We first search for posts that contain the tag \emph{smart contract}, and then iteratively extend our search with new tags, finally stopping at the following list of tags:
\emph{solidity},
\emph{smart contract},
\emph{smart contracts},
\emph{vyper},
\emph{metamask},
\emph{truffle},
\emph{erc20},
\emph{web3}.
Again, we manually check a random sample to confirm that the collected post are indeed related to smart contracts.
In total, we collect 4,045 unique posts from 2,165 authors, which have been posted on Medium between \Aron{Are there no matching posts before Jan 2014 or do we not collect them?} \Shanto{Prof., I got only one that was published maybe in 2012/2013. That is why skipped that one.} January 2014 and November 24, 2019.


\subsection{Methodology}
\label{sec:method}


\subsubsection{Statistical Analysis}
%
First, we analyze various statistics of smart contract related posts and the posting users  
from Stack Exchange and Medium.
Statistics for posts include the rate of new posts over time, the distributions of tags, number of answers,  etc.,
while statistics for users include the distribution of tags in all of their posts.
For the Stack Exchange dataset, we also compare smart contract related posts to posts about other subjects on Stack Overflow.
\SPACE{
\Ayman{Need to change this workflow}
\begin{figure}
\vspace{-1.5em}
    \centering
    \includegraphics[width=\textwidth]{images/Process_Workflow.pdf}
    \caption{Workflow of our study.}
    \label{fig:Process_Workflow}
\end{figure}
}
\subsubsection{Textual Data Analysis}
Next, we preprocess the data to prepare the posts for text analysis. 
%
First, for each Stack Exchange post, we combine the title, question, answers, comments, and tags together; for each Medium post, we combine the title, text, and tags together.
Second, we remove HTML tags and code snippets from all posts.
After this step, we search for occurrences of certain keywords in the posts, such as mentions of common security issues and tools.

\ClearPage
\section{Results}
\label{sec:results}


\subsection{Discussion Trends (Q1)}

\begin{figure}[ht!]
\vspace{-2em}
\pgfplotstableread[col sep=comma]{Data/Trend_with_AllPosts_JAVA_Python_JS.csv}\table

    \centering
    \begin{tikzpicture}
    \begin{axis}[
    ybar,
    ymin = 0,
    xmin=-1,
    xmax={54},
    scale only axis,
    axis y line*=left,
    width=0.75\textwidth,
    height=3.5cm,
    xtick = data,
    xticklabels from table={\table}{Months},
    ylabel style = {align = center},
    ylabel=Smart Contract\\Related Questions,
     legend style={at={(0,1)},anchor=south west,font=\tiny},
     xtick pos=left,
    ]
    \addplot [BlueBars] table [x expr=\coordindex, y=All_Posts] {\table};
    \end{axis}

    \begin{axis}[
    scale only axis,
    axis y line*=right,
    axis x line=none,
    xtick = data,
    width=0.75\textwidth,
    height=3.5cm,
    ymin=0,
    xmin=-1,
    xmax={54},
    ymax = 250000,
    ylabel style = {align = center},
    ylabel=Other Questions,
    legend style={at={(1,1)},anchor=south east,font=\tiny}
    ]
    \addplot [black, mark=none, ultra thick] table [x expr=\coordindex, y=All_General_Posts_SO] {\table};
     
     \addplot[dotted, blue, ultra thick] table [x expr=\coordindex, y=Java_Posts_SO] {\table};
     
     \addplot[dashed, red, ultra thick] table [x expr=\coordindex, y=Python_Posts_SO] {\table};
     
     \addplot[dashdotted, ForestGreen, ultra thick] table [x expr=\coordindex, y=JS_Posts_SO] {\table};

    \end{axis}
\end{tikzpicture}
            
    \caption{Number of smart contract related questions (vertical bars) posted on Stack Exchange, number of all questions (black line) and Java (\textcolor{blue}{dotted blue}), Python (\textcolor{red}{dashed red}), and JavaScript (\textcolor{ForestGreen}{dash-dotted green}) related questions posted on Stack Overflow each month. Please note the different scales.}
    \label{fig:SO_NumberOfPosts_Comparison}
\end{figure}
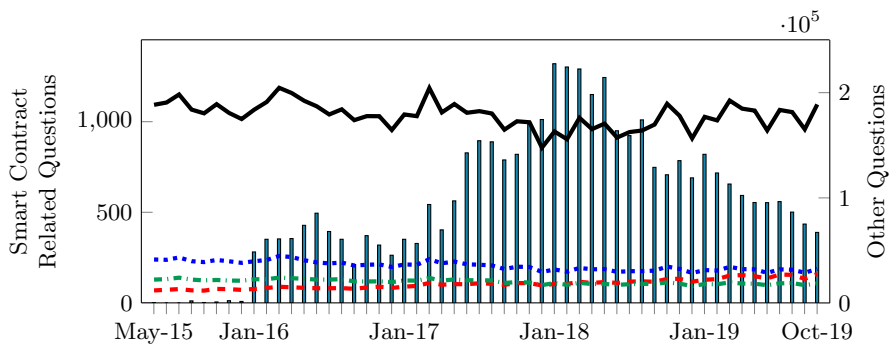

\begin{figure}[h!]
    \centering
\pgfplotstableread[col sep=comma,]{Data/medium_post_count.csv}\datatable
\begin{tikzpicture}
\begin{axis}[
    xtick pos=left,
    ybar,
    bar width=.1cm,
    ymin = 0,
    xmin={-1},
    xmax={59},
    width=0.87\textwidth,
    height=4cm,
    xtick=data,
    xticklabels from table={\datatable}{month},
    ylabel={Number of Posts}]
    \addplot [BlueBars] table [x expr=\coordindex, y={count}]{\datatable};
\end{axis}
\end{tikzpicture}
\caption{Number of smart contract related posts on Medium each month.}
\label{fig:medium_post_freq}
\end{figure}
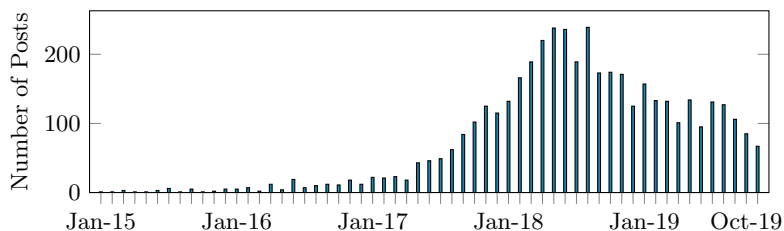

We begin our analysis by comparing trends in posts about smart contracts with trends in posts about other technologies (e.g., Java and Python). 
Specifically, we study how interest in smart contracts (measured as the number of new posts) has evolved over time and how active smart contract related discussions are (measured using distributions of answers and comments), showing significant differences compared to other technologies.


Figure~\ref{fig:SO_NumberOfPosts_Comparison} compares the number of questions related to smart contracts (vertical bars) posted on Stack Exchange with the total number of questions (\textcolor{black}{black line}) posted on Stack Overflow each month.
For the sake of comparison, the figure also shows numbers of questions about other, more mature technologies, namely Java (\textcolor{blue}{dotted blue}), Python (\textcolor{red}{dashed red}), and JavaScript (\textcolor{ForestGreen}{dash-dotted green}).
The first smart contract related questions were posted in May 2015, but users did not start posting in significant numbers until Ethereum Stack Exchange launched in January 2016.
From 2017 to early 2018, there is a clear upward trend; however, the rate of new questions has been steadily declining since then, which suggests that interest in smart contracts on Stack Exchange peaked in early 2018. 
Meanwhile,  the overall rate of new questions on Stack Overflow has remained steady since 2015.
Similarly, the rates of new questions about Java, Python, and JavaScript have remained relatively steady, with only slightly increasing (Python) and decreasing (Java) trends.
%
These results suggest that the significant fluctuations observed in smart contract related questions are not due to the varying popularity of Stack Overflow.
Finally, Figure~\ref{fig:medium_post_freq} shows the number of new Medium posts related to smart contracts in each month.
Again, we observe a clear upward trend from 2017 to early 2018, peaking in the first half of 2018, and a steady decrease since then. 

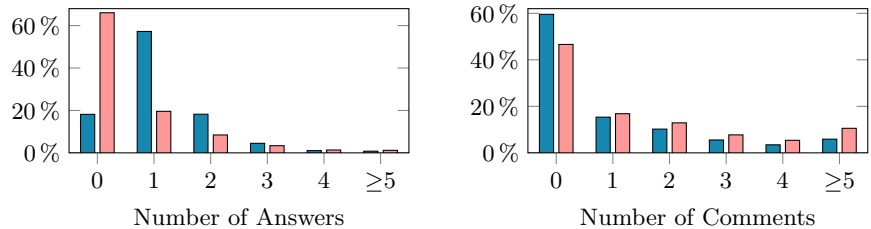
\begin{figure}[th]
\pgfplotstableread[col sep = comma]{Data/AnswerCommentCount.csv}\table
\centering
\begin{subfigure}[b]{.5\linewidth}
    \centering
    \begin{tikzpicture}
    \begin{axis}[
            SmallBarPlot,
            yticklabel=\pgfmathprintnumber{\tick}\,$\%$,
            ymax = 68,
            height = 3.5cm, 
            xticklabels={0, 1, 2, 3, 4, {$\geq$5}}, 
            ylabel style = {align = center},
            xlabel=Number of Answers
        ]
        \addplot [BlueBars] table [x expr=\coordindex, y=PercentageAnswer] {\table};
        \addlegendentry{Smart Contracts}
        \addplot [RedBars] table [x expr=\coordindex, y=PercentageAnswerGeneral] {\table};
        \addlegendentry{General StackOverflow}
        \legend{}
    \end{axis}
    \end{tikzpicture}
\end{subfigure}%
\begin{subfigure}[b]{.5\linewidth}
     \centering
    \begin{tikzpicture}
    \begin{axis}[
            SmallBarPlot,
            yticklabel=\pgfmathprintnumber{\tick}\,$\%$,
            height = 3.5cm, 
            ymax = 62,
            xticklabels={0, 1, 2, 3, 4, {$\geq$5}}, 
            ylabel style = {align = center},
            xlabel= Number of Comments
        ]
        \addplot [BlueBars] table [x expr=\coordindex, y=PercentageComments] {\table};
        \addlegendentry{Smart Contracts}
        \addplot [RedBars] table [x expr=\coordindex, y=PercentageCommentsGeneral] {\table};
        \addlegendentry{General StackOverflow}
        \legend{}
    \end{axis}
    \end{tikzpicture}
\end{subfigure}
\caption{Number of answers and comments received by smart contract related questions \LegendBox{MidnightBlue}{blue} and by other questions \legendbox{red} on Stack Exchange.}
\label{fig:SO_Answer_Comment_Count_Comparison}
\end{figure}

Further, we can see a very similar trend in the price of Ethereum (ETH)~\footnote{\url{www.coinbase.com/price/ethereum}} over the past years: ETH reached its highest value on January 12, 2018 \cite{ethereumprice:online} and has been mostly declining since then. 
The close similarity between Figures~\ref{fig:SO_NumberOfPosts_Comparison} and~\ref{fig:medium_post_freq} as well as the decreasing price trend of ETH  suggest that our observations are robust in the sense that they are not artifacts of our data sources or our analysis; rather, the trends that we observe may be signs of declining developer interest in smart contracts. 

To gain insight into the level of interactions in the smart contract developer community, we analyze the distributions of answers and comments in smart contract related posts.
Figure~\ref{fig:SO_Answer_Comment_Count_Comparison} shows the  number of answers and comments received by smart contract related questions \LegendBox{MidnightBlue}{blue} on Ethereum Stack Exchange and Stack Overflow and by other questions \LegendBox{red}{red} on Stack Overflow. 
We observe that 82\% of smart contract related questions have at least one answer.
This ratio is very high compared to other questions, of which less than 34\% have at least one answer.
We speculate that this difference could be explained by smart contract related questions being simpler and easier to answer, asking mostly about basic issues; or it could be explained by the community of smart contract developers being more active.
However, we also observe that few smart contract related questions receive more than one answer, and very few receive five or more, especially in comparison with other questions.
This suggests that the more likely explanation is that smart contract related questions are indeed simpler since developers rarely post improved or conflicting answers.
We also observe that smart contract related questions tend to receive fewer comments than other questions, and receiving five or more comments is very rare.
In other words, smart contract related questions rarely spark lengthy debates or discussions, which again might suggest that  questions pertain to simpler issues.

Finally, we study what topics are at the center of smart contract related discussions.
Posts from both Stack Exchange and Medium have tags to identify the topic of discussion. Although these tags do not necessarily capture the exact topic of discussion, they can indicate what technologies, issues, etc. are discussed.
Table~\ref{tab:Popular_TagFrequency} lists ten tags that are most frequently used in smart contract related posts on each site. 
For Stack Exchange, the table lists the average score\footnote{Score is the difference between the number of upvotes and downvotes for a post.} and the average number of views, answers, and comments received by questions with each tag.
For Medium, it lists the average number of responses, claps, and number of voters for each tag. 
The list is dominated by a few  smart contract technologies, such as
\emph{Solidity} (high-level language for smart contracts), \emph{Go-Ethereum} (official implementation of Ethereum),  \emph{web3js} (JavaScript library for Ethereum), and \emph{Truffle} (development environment for smart contracts).
This may suggest the existence of a monoculture: most developers might be familiar with only a small set of technologies.

\newcolumntype{M}[1]{>{\centering\arraybackslash}m{#1}}
\begin{table}[t!]
\caption{Most Frequent Tags in Smart Contract Related Posts}
\label{tab:Popular_TagFrequency}

     \centering
     \resizebox{\columnwidth}{!}{%
    \begin{tabular}{|M{1.9cm}|M{1cm}|M{1cm}|M{1.2cm}|M{1cm}|M{1cm}||M{1.7cm}|M{1cm}|M{1cm}|M{1cm}|M{1cm}|}
    \hhline{------||-----}
    \multicolumn{6}{|c||}{\textbf{Stack Exchange}} & \multicolumn{5}{c|}{\textbf{Medium}}\\
    \hhline{------||-----}
    \multirow{2}{*}{\textbf{Tag}} & \multirow{2}{*}{\textbf{Num.}} & \multicolumn{4}{c||}{\textbf{Average}} &
    \multirow{2}{*}{\textbf{Tag}} & \multirow{2}{*}{\textbf{Num.}} & \multicolumn{3}{c|}{\textbf{Average}}\\
    \hhline{~~----||~~---}
     &  & \textbf{Score} & \textbf{View} & \textbf{Ans.} & \textbf{Com.} &  &  & \textbf{Resp.} & \textbf{Clap} & \textbf{Voter} \\ \hhline{------||-----}\hhline{------||-----}

Solidity & 9323 & 0.48& 752 & 1.2& 1.14 & Ethereum  &  2643 &  2.37 &  388 &  37.10 \\
 \hhline{------||-----}
 Go-Ethereum & 4946 & 0.55& 1047 & 1.09& 1.19 & Blockchain & 2585 &  2.06 &  423 & 35.91 \\
 \hhline{------||-----}
 web3js & 3948 & 0.48& 880 & 1.16& 1.37 & Smart Contracts & 1274 & 1.68 &  311 &  32.03 \\
 \hhline{------||-----}
 Contract-development & 2973 & 0.70 & 845 & 1.29& 1.04 & Solidity & 907 &  1.62 &  290 &  29.26 \\
 \hhline{------||-----}
 Blockchain & 2539 & 0.88& 1232 & 1.53& 1.37 & Crypto- currency & 659 & 2.48 &  577 &  41.32\\
 \hhline{------||-----} 
 Ethereum & 2530 & 1.55& 3023 & 3.73& 3.94 & Security & 476 &  0.81 &  194 &  16.50  \\
 \hhline{------||-----}
 Truffle & 2430 & 0.40 & 750 & 1.28& 1.46 & ERC20 & 467 &  3.63 &  836 &  54.46  \\
 \hhline{------||-----} 
 Transactions & 1743 & 0.94 & 1382 & 1.31& 1.1 & Web3 & 401 &  1.75 &  429 &  41.03\\
 \hhline{------||-----}
 Remix & 1642 & 0.29& 593 & 1.15& 1.34 & Bitcoin & 369 &  5.04 &  730 &  74.07 \\
 \hhline{------||-----}
 Contract-design & 1522 & 1.12 & 873 & 1.34& 0.92 & MetaMask & 296 &  0.76 &  216 &  16.97\\

 \hhline{------||-----}
\end{tabular}
}
\end{table}

\subsection{Security Issues and Tools (Q2)}
\label{sec:security}
Next, we focus on discussions related to security.
Our goal is to gauge the smart contract developers' level of concern and awareness about various security issues and tools. 
To this end, we search for posts related to common security issues and tools, using the numbers of related posts as indicators for concern about security and for awareness about tools.

\subsubsection{Security Issues}

\begin{table}[ht]
\centering
\caption{Posts Mentioning Common Security Issues}
\label{tab:SO_Security_Issue_Mentions}
\resizebox{\columnwidth}{!}{%
\begin{tabular}{|c||c|c||c|c|}
\hhline{-||--||--}
\multirow{2}{*}{\textbf{Security Issues}}& \multicolumn{2}{c||}{\textbf{Stack Exchange}}  & \multicolumn{2}{c|}{\textbf{Medium}} \\ \hhline{~||--||--}

& \textbf{Number} & \textbf{Percentage} & \textbf{Number} & \textbf{Percentage}\\

\hhline{-||--||--}
\hhline{-||--||--}
Re-Entrancy & 126 & 0.41\% & 164 & 4.05\%\\
\hhline{-||--||--}
Denial of Service & 95 & 0.31\% & 111 & 2.74\%\\
\hhline{-||--||--}
Race Condition & 35 & 0.11\% & 34 & 0.84\%\\
\hhline{-||--||--}
Integer Overflow & 16 & 0.05\% & 95 & 2.35\% \\
\hhline{-||--||--}
Transaction-Ordering Dependence & 4 & 0.01\% & 66 & 1.63\%\\
\hhline{-||--||--}
Timestamp Dependence &  4 & 0.01\% & 49 & 1.21\%\\
\hhline{-||--||--}
Integer Underflow & 2 & 0.007\%& 12 & 0.30\%\\
\hhline{-||--||--}
\end{tabular}
}
\end{table}


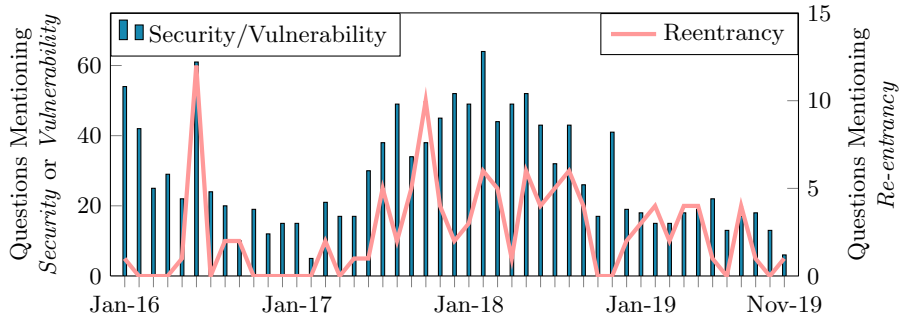
\begin{figure}[ht]
\pgfplotstableread[col sep=comma]{Data/Trends_vulnerability.csv}\table

    \centering
    \begin{tikzpicture}
    \begin{axis}[
    xtick pos=left,
    ybar,
    ymin = 0,
    xmin={-1},
    xmax={47},
    scale only axis,
    axis y line*=left,
    width=0.75\textwidth,
    height=3.5cm,
    xtick = data,
    ymax = 75,
    xticklabels from table={\table}{Months},
    ylabel style={align=center},
    ylabel=Questions Mentioning\\ \emph{Security} or \emph{Vulnerability},
    legend style={at={(0,1)},anchor=north west}
    ]
    \addplot [BlueBars] table [x expr=\coordindex, y=Sec_Vul] {\table};
    \addlegendentry{Security/Vulnerability}
    \end{axis}

    \begin{axis}[
    scale only axis,
    axis y line*=right,
    axis x line=none,
    xtick = data,
    ymax = 15,
    xmin={-1},
    xmax={47},
    width=0.75\textwidth,
    height=3.5cm,
    ymin=0,
    ylabel style={align=center},
    ylabel=Questions Mentioning\\\emph{Re-entrancy},
    legend style={at={(1,1)},anchor=north east}
    ]
    \addplot [red!40, mark=none, ultra thick] table [x expr=\coordindex, y=Reentrancy_List] {\table};
     \addlegendentry{Reentrancy}
    \end{axis}
\end{tikzpicture}
            
    \caption{Number of smart contract posts per month on Stack Exchange mentioning \emph{security} or \emph{vulnerability} and \emph{re-entrancy}. Please note the different scales.}
    \label{fig:SO_Trend_for_Reentracy_Security_Vulnerability}
\end{figure}

To gauge how concerned smart contract developers are about security,
we first search for mentions of \emph{security} and \emph{vulnerability} in smart contract related posts.
We search in the preprocessed texts of the posts, which include tags, titles, comments, etc.  
by considering all common variations of these terms (e.g., \emph{vulnerabilities} and \emph{vulnerable}).
On Stack Exchange, we find 1,211 and 236 posts that mention \emph{security} and \emph{vulnerability}, respectively, which constitute only 3.9\% and 0.77\% of all smart contract related posts.  
On Medium, we find 1,429 and 470 posts that mention \emph{security} and \emph{vulnerability}, respectively, which constitute 32\% and 11\% of all smart contract related posts.
Based on these findings, we speculate that security awareness on Stack Exchange is rather low, while it is comparatively high on Medium.
Unfortunately, many developers use Stack Exchange as their primary source of information~\cite{Barua2014}.

Next, we consider specific types of vulnerabilities.
Based on prior surveys of smart contract vulnerabilities~\cite{luu2016making,atzei2017survey,li2017survey,zhu2018research,chen2019survey}, we establish the following list of common issues to search for: \emph{re-entrancy}, 
\emph{timestamp dependence}, 
\emph{transaction-ordering dependence},
\emph{integer overflow},  \emph{integer underflow},
\emph{race condition}, and
\emph{denial of service}.
Again, we search for mentions of these issues in the preprocessed posts by considering all common variations (e.g., \emph{DoS}, \emph{dependence} and \emph{dependency}).
Table~\ref{tab:SO_Security_Issue_Mentions} shows the number of smart contract related Stack Exchange and Medium posts that mention these issues.
We find that Stack Exchange not only suffers from generally low security concern, but discussions are also restricted to only a few issues, such as \emph{re-entrancy}; meanwhile, Medium posts discuss a broader range of issues.
To explain  Stack Exchange users' fascination with the \emph{re-entrancy} vulnerability, consider Figure~\ref{fig:SO_Trend_for_Reentracy_Security_Vulnerability}, which shows the number of new posts mentioning \emph{re-entrancy} for each month.
There is a significant peak in 2016 June, which is when one of the most famous Ethereum security incidents happened, the so-called ``DAO attack,'' which exploited a \emph{re-entrancy} vulnerability~\cite{finley2016million}.
A significant number of security discussions on Stack Exchange seem to be driven by this incident.
Also note that Figure~\ref{fig:SO_Trend_for_Reentracy_Security_Vulnerability} shows relatively high interest in security back in 2016.
However, while the number of smart contract related posts on Stack Exchange rapidly rose in 2017, interest in security rather declined.





\subsubsection{Security Tools, Frameworks, and Design Patterns}
\newcolumntype{M}[1]{>{\centering\arraybackslash}m{#1}}

\begin{table}[t!]
\caption{Number of Posts Mentioning Various Security  Tools and Patterns}
\label{tab:SO_Tools_Framework_Mentions}
\centering
\begin{tabular}{|M{2.8cm}|M{1.7cm}|c||M{2.4cm}|M{1.7cm}|c|}
\hhline{---||---}
\textbf{Tools and Pattern}& \textbf{Stack Exchange} & \textbf{Medium} & \textbf{Tools and Pattern}& \textbf{Stack Exchange} & \textbf{Medium}\\

\hhline{---||---} Mythril & 12 & 98 & solcheck & 2 & 5\\
\hhline{---||---} Oyente & 10 & 64 & Maian & 2 & 3\\ 
\hhline{---||---} Smartcheck & 4 & 57 & Octopus & 0 & 3\\
\hhline{---||---} Securify & 6 & 46 & teEther & 6 & 2 \\
\hhline{---||---} Solhint & 8 & 39 & Vandal & 2 & 2\\
\hhline{---||---} Ethlint/Solium & 6 & 36 & EthIR & 2 & 2 \\
\hhline{---||---} scompile & 0 & 33 & SASC & 1 & 2\\
\hhline{---||---} Checks-Effects-Interactions & 22 & 17 & VeriSolid & 0 & 2\\
\hhline{---||---} Manticore & 3 & 16 & Zeus & 1 & 1\\
\hhline{---||---} Slither & 0 & 10 & Rattle & 1 & 1\\
\hhline{---||---} solgraph & 2 & 7 & ContractFuzzer & 1 & 1\\
\hhline{---||---} solint & 2 & 6 & SonarSolidity & 3 & 0\\ 
\hhline{---||---} Surya (Sūrya) & 0 & 6 & echidna & 1 & 0 \\
\hhline{---||---}

\end{tabular}
\end{table}

%
%
\TODO{Revise this sentence}
We complement our results on security issues by studying the smart contract developers' awareness of security tools (e.g., which tools  they ask about or suggest in answers). 
We compile a comprehensive list of security tools based on relevant evaluation and survey papers (e.g.,~\cite{di2019survey,kirillov2019evaluation,chen2019survey,liu2019survey,harz2018towards,parizi2018empirical})\extended{ and other sources (e.g.,~\cite{consensys2019security})},
%
and search for mentions of the following (in alphabetical order): 
ContractFuzzer\extended{~\cite{jiang2018contractfuzzer}},
ContractLarva\extended{~\cite{ellul2018runtime}}, 
echidna\extended{\footnote{\url{github.com/crytic/echidna}}},
EtherTrust\extended{~\cite{grishchenko2018ethertrust}},
EthIR, 
Ethlint (formerly known as Solium)\extended{\footnote{\url{www.ethlint.com}}},
FSolidM\extended{~\cite{mavridou2018designing}}, 
MAIAN\extended{~\cite{nikolic2018finding}},
Manticore\extended{~\cite{mossberg2019manticore}},
Mythril (as well as the service MythX and the client Mythos)\extended{~\cite{mueller2018smashing}}, 
Octopus\extended{\footnote{\url{github.com/quoscient/octopus}}},
Osiris\extended{~\cite{torres2018osiris}},
Oyente\extended{~\cite{luu2016making}}, 
Rattle\extended{~\cite{stortz2018rattle}},
ReGuard\extended{~\cite{liu2018reguard}},
SASC\extended{~\cite{zhou2018security}},
sCompile\extended{~\cite{chang2018scompile}}, 
Securify\extended{~\cite{tsankov2018securify}}, 
Slither\extended{~\cite{feist2019slither}},
SmartAnvil\extended{~\cite{ducasse2019smartanvil}},
SmartCheck\extended{~\cite{tikhomirov2018smartcheck}}, 
solcheck\extended{\footnote{\url{github.com/federicobond/solcheck}}}, 
solgraph\extended{\footnote{\url{github.com/raineorshine/solgraph}}}, 
solint\extended{\footnote{\url{github.com/SilentCicero/solint}}}, 
Solhint\extended{\footnote{\url{protofire.github.io/solhint}}}, 
SonarSolidity\extended{\footnote{\url{github.com/sagap/sonar-solidity}}}, 
Sūrya (also spelled as Surya)\extended{\footnote{\url{github.com/ConsenSys/surya}}}, 
teEther\extended{~\cite{krupp2018teether}},
Vandal\extended{~\cite{brent2018vandal}}, 
\TODO{VeriSol from Microsoft?}
VeriSolid\extended{~\cite{mavridou2019verisolid}},
VerX\extended{~\cite{permenev2019verx}},
VULTRON\extended{~\cite{wang2019vultron}}, 
Zeus\extended{~\cite{kalra2018zeus}}.
Note that our goal is not to evaluate or compare the technical quality of these tools and frameworks (for that we refer the reader to surveys, e.g.,~\cite{parizi2018empirical}); we are only interested in whether they are discussed by developers.  
We also search for mentions of the \emph{checks-effects-interactions} design pattern---considering again variations in spelling---which is meant to prevent the re-entrancy vulnerability~\cite{solidityCEIpattern}.

Table \ref{tab:SO_Tools_Framework_Mentions} shows the number of smart contract related posts on Stack Exchange and Medium that mention the above tools.
We again find low awareness on Stack Exchange: Mythril and Oyente are mentioned by only 12 and 10 posts, and other tools are mentioned by even fewer.
However, we do find 22 posts that mention the \emph{checks-effects-interactions} pattern, which is most likely due to interest in the \emph{re-entrancy} vulnerability (see Table~\ref{tab:SO_Security_Issue_Mentions}).
Similarly, we again find higher awareness on Medium: there are 7 tools that are mentioned at least 33 times, with Mythril being mentioned the most.

\subsubsection{Co-Occurrence of Security Issues and Tools}

\newcolumntype{M}[1]{>{\centering\arraybackslash}m{#1}}

\begin{table}[t!]
\caption{Co-Occurrence of Security Issues and Tools in Posts}
\centering
\resizebox{\columnwidth}{!}{%
\begin{tabular}{|M{2.2cm}||M{1.5cm}|M{1.4cm}||M{1.5cm}|M{1.4cm}||M{1.5cm}|M{1.4cm}||M{1.5cm}|M{1.4cm}|}
\hhline{-||--||--||--||--}
 & \multicolumn{2}{M{2.9cm}||}{\textbf{Security /Vulnerability}} & \multicolumn{2}{M{2.9cm}||}{\textbf{Re-Entrancy}} & \multicolumn{2}{M{2.9cm}||}{\textbf{Timestamp Dependency}} & \multicolumn{2}{M{2.9cm}|}{\textbf{Transaction Ordering Dependency}} \\ \hhline{-||--||--||--||--}
\textbf{Security Tools} & \textbf{Stack Overflow} & \textbf{Medium}  & \textbf{Stack Overflow} & \textbf{Medium}  & \textbf{Stack Overflow}  & \textbf{Medium}  &  \textbf{Stack Overflow}  & \textbf{Medium} \\ \hhline{-||--||--||--||--}\hhline{-||--||--||--||--}
Mythril & 6 & 95  &  2  &  36  & 2  &  19  &  2  &  17   \\ \hhline{-||--||--||--||--}
Oyente &  6  &  63  &  2  &  37  &  1   &   26  & 1  &  3 \\ \hhline{-||--||--||--||--}
Smartcheck &  3  &  57  & 0    &  45  &  1   & 34  &  1  & 2 \\ \hhline{-||--||--||--||--}
Securify &  4  &  45   &  1  & 26   &  1  &  20   &  1  &  3  \\ \hhline{-||--||--||--||--}
Solhint &  2   &  38  &  1   &  34    & 1   &  26  &  1   &  1  \\ \hhline{-||--||--||--||--}
Ethlint/Solium &  4 &  28   &  1  &  10  & 1  &  6  &  1  & 3  \\ \hhline{-||--||--||--||--}
Manticore &  3  & 16  &  1  &  5  &  1   &  1   &  1  &  0  \\ \hhline{-||--||--||--||--}
Slither & 0  &  10  &  0  &  7   &  0  &  3  &  0  & 2  \\ \hhline{-||--||--||--||--}
solgraph &  2  &  7   & 1    & 4 & 1 &  1  &  1  & 0 \\ \hhline{-||--||--||--||--}
Surya & 0  &  6  &  0  &  2   &  0  &  1  &  0  & 0  \\ \hhline{-||--||--||--||--}
solint &  2  &  5 &  1   &  1  &  1  &  0  &  1  & 0 \\ \hhline{-||--||--||--||--}
Solcheck  &  2  &   4  & 1  &  2  & 1  &  1  & 1  &  0 \\ \hhline{-||--||--||--||--}
Maian & 0  &  3  &  0  &  2   &  0  &  1  &  0  & 1  \\ \hhline{-||--||--||--||--}

SASC & 0  &  2  &  0  &  1   &  0  &  1  &  0  & 1  \\ \hhline{-||--||--||--||--}
VeriSolid & 0  &  2  &  0  &  1   &  0  &  0  &  0  & 0  \\ \hhline{-||--||--||--||--}
Vandal & 2 &  2  &  0  &  1   &  0  &  0  &  0  & 0  \\ \hhline{-||--||--||--||--}


teEther & 2 &  1  &  0  &  0   &  0  &  0  &  0  & 0  \\ \hhline{-||--||--||--||--}
EthIR & 2 &  1  &  0  &  1   &  0  &  0  &  0  & 0  \\ \hhline{-||--||--||--||--}
\end{tabular}
}
\label{tab:Co_Occurance}
\end{table}

\newcolumntype{M}[1]{>{\centering\arraybackslash}m{#1}}

\begin{table}[ht]
\caption{Other Tags used by Smart Contract Developers on SO \& Medium}
\label{tab:DifferentTagsUsed_by_User}
\resizebox{\columnwidth}{!}{%
\begin{tabular}{|M{2.1cm}|M{1cm}||M{2.5cm}|M{1cm}|M{2.3cm}|M{1cm}|M{2.5cm}|M{1cm}|}
\hhline{--||------}
\multicolumn{2}{|M{3.1cm}||}{\multirow{2}{*}{\textbf{Stack Overflow}}} & \multicolumn{6}{M{10.3cm}|}{\textbf{Medium}} \\ \hhline{~~||------}\hhline{~~||------}
\multicolumn{2}{|M{3.1cm}||}{\textbf{}} &
\multicolumn{2}{M{3.5cm}|}{\textbf{Blockchain}} & \multicolumn{2}{M{3.3cm}|}{\textbf{Technical (Other)}} & \multicolumn{2}{M{3.5cm}|}{\textbf{Non-Technical}} \\ \hhline{--||------}
 \textbf{Tag}  & \textbf{Freq.} & \textbf{Tag}  & \textbf{Freq.} & \textbf{Tag }& \textbf{Freq.} & \textbf{Tag} & \textbf{Freq.} \\ \hhline{--||------}\hhline{--||------}
 
\textcolor{blue}{jQuery} & 8787 & Blockchain  & 17877  & Technology & 2167 & Startup & 1897 \\ \hhline{--||------}
HTML & 6503 & Cryptocurrency & 8957  & Artificial Intelligence & 1331  & Investing & 958 \\ \hhline{--||------}
CSS & 5657 & Bitcoin & 5013 & Fintech & 1232 & Finance & 820 \\ \hhline{--||------}
\textcolor{blue}{Node.js} & 5040 & Crypto & 2511 & IoT & 697 & Business & 786 \\ \hhline{--||------}
.NET & 4247 & ICO & 2256 & Programming & 646 & Entrepreneurship & 582 \\ \hhline{--||------}
Android & 3739 & Security & 630 & \textcolor{blue}{JavaScript} & 635 & Exchange  & 527 \\ \hhline{--||------}
Objective-C & 3727 & Cryptocurrency Investment  & 620  & Machine Learning & 479 & Marketing & 493 \\ \hhline{--||------}
MySQL & 3330 & Token Sale &  616 & Software Development & 376  & Innovation  & 488 \\ \hhline{--||------}
Ruby & 3281 & Decentralization & 525 & Privacy & 342 & News & 467 \\ \hhline{--||------}
\textcolor{blue}{JSON}  & 3231 & Tokenization & 220 & Data & 329 & Travel & 428\\ \hhline{--||------}
\end{tabular}
}
\end{table}

Finally, we investigate if users recommend these tools against certain vulnerabilities and if they are aware of which vulnerabilities these tools address.
To this end, we study which security issues and tools are mentioned together.
Table~\ref{tab:Co_Occurance} shows the number of posts on  Stack Exchange and Medium that mention various pairs of security issues and tools (focusing on pairs mentioned by the most posts, omitting less frequent pairs).
Again, we find low awareness on Stack Exchange: Mythril and Oyente are each mentioned only in 6 posts that also mention \emph{security} or \emph{vulnerability}, which means that these tools are suggested for security issues less than 0.5\% of the time; other tools are mentioned even fewer times. These tools are not mentioned even in conjunction with vulnerabilities that they address (see, e.g., \emph{re-entrancy}).
On the other hand, we find much higher awareness on Medium, as security issues and tools are often mentioned together.

\SPACE{
\ClearPage
\subsection{Discussion Topics (Q2)}
\label{sec:topics}
\subsubsection{Frequency and Popularity of Tags}
First, we study which tags are used most frequently in smart contract related posts\Amin{After some point in the paper, you can use ``the posts'' instead of ``smart contract related posts''.}\Aron{Or ``SC posts'' if we want to keep it short. Sometimes, we draw comparison with other posts, so it would be helpful to always make it clear what posts we mean.} on Stack Exchange and Medium.
Although tags do not necessarily capture the exact topic of discussion, they can indicate what technologies and concepts are involved.
Table~\ref{tab:Popular_TagFrequency} lists ten tags that are most frequently used in smart contract related questions on both Stack Exchange and Medium. We also calculate average scores \footnote{measures the difference between the number of upvotes and downvotes for a post}, views, answers, and comments received by questions with that tag.
For medium, we present the average number of responses, claps, and number of voters for each tag. 
\TODO{We may need to revise this based on the corrected Medium results!}
The list is dominated by a few  smart contract technologies, such as
\emph{solidity} (high-level language for smart contracts), \emph{go-ethereum} (official implementation of Ethereum),  \emph{web3js} (JavaScript library for Ethereum), and \emph{truffle} (development environment for smart contracts).
This suggests the existence of a monoculture: most developers are familiar with a small set of technologies.
It is also interesting to note that by all metrics, questions tagged with \emph{ethereum} and \emph{ether} receive much more attention than others. For Medium, we see that \emph{Solidity}, and \emph{Web3} again appear among the top tags. It is also interesting to note that \emph{Security} also appears on the list.






\begin{table}[t]
\centering
\caption{Automated Topic Modeling on Stack Exchange Posts}
\label{tab:Automated_Topic_Analysis}
\scalebox{0.9}{\begin{tabular}{|c|l|}
\hline
\textbf{Topic Name} & \textbf{Top Words in Topic} \\
\hline
\hline
\textbf{Ethereum basics} & 
\Large{ethereum}, \large{blockchain, contract}, \normalsize{use}, \small{data}, \scriptsize{would, smart, like}, \tiny{need, question} \\
\hline
Gas and transaction costs &
\scriptsize gas, block, transaction, time, cost, miner, limit, price, number, ethereum \\
\hline
Tokens and ether &
\scriptsize token, address, contract, account, ether, wallet, key, transfer, send, eth \\
\hline
Development environments &
\scriptsize error, truffle, using, code, web3, version, test, contract, remix, deploy \\
\hline
Events and logging &
\scriptsize transaction, event, hash, get, block, log, nonce, send, signature, message \\
\hline
GPU mining &
\scriptsize card, gpu, rig, constantinople, admin, checksum, gpus, hardware, 1x, graphic \\
\hline
Private networks &
\scriptsize node, geth, network, private, account, using, file, running, ethereum, run \\
\hline
Solidity function calls &
\scriptsize contract, function, call, code, solidity, value, address, return, array, smart \\
\hline
\end{tabular}}
\end{table}

\vspace{-0.5em}
\subsubsection{Topic Analysis}
To find the key topics that smart contract developers discuss on Stack Exchange, we apply automated topic modeling to the smart contract related posts (see Section~\ref{sec:method}).
Table \ref{tab:Automated_Topic_Analysis} shows the results of this topic analysis as groups of words (right column) that are frequently mentioned together.
Based on these words, we gave an intuitive name to each topic (left column).
These range from topics that are strictly related to smart contract development, such as \emph{Solidity function calls}, to more general blockchain topics, such as \emph{GPU mining}.




\ClearPage


}

\subsection{Developers' Background and Interests (Q3)}
\label{sec:community}

\TODO{Revise!}
For many developers, it is easier to adopt new tools, languages, and platforms that resemble ones with which they are already familiar. 
Hence, adoption of new technologies 
can hinge on the developers' technological background.
To discover with which technologies smart contract developers are familiar, we study what tags they use in posts that are \emph{not} related to smart contracts.

For each smart contract developer, we retrieve all of the developer's posts (i.e., questions and answers, or blog posts) that are \emph{not} related to smart contracts, collecting a total of 1,250,325 posts from Stack Overflow and 44,684 posts from Medium. 
Table \ref{tab:DifferentTagsUsed_by_User} lists the 10 most frequently used tags in the smart contract developers' Stack Overflow posts.
The most frequent tags are all related to web development (\emph{jQuery}, \emph{HTML}, \emph{CSS}, \emph{Node.js}).
\Amin{can we be more specific and say that are more background in web development?}\Aron{Yes, JavaScript is particularly interesting, see end of the next paragraph}
Other popular tags correspond to major platforms (\emph{.NET}, \emph{Android}).
Table \ref{tab:DifferentTagsUsed_by_User} lists the most frequent tags from the smart contract developers' Medium posts in three categories: blockchain related, other technical, and non-technical (i.e., everything else). Note that since Medium is a generic blog site, there are many non-technical posts (e.g., tagged with \emph{Business} or \emph{Travel}).
Unsurprisingly, the most popular tags are related to blockchains and cryptocurrencies.
Other technical terms are led by the area of \emph{Artificial Intelligence} and \emph{Machine Learning}, and by tags related to software development (e.g., \emph{Programming} and \emph{JavaScript}). 
The most frequent non-technical terms are related to entrepreneurship (\emph{Startup}, \emph{Finance}, \emph{Business}, \emph{Investing}, etc.).

On both sites, we observe that a significant number of posts are related to JavaScript (highlighted in \textcolor{blue}{blue} in Table~\ref{tab:DifferentTagsUsed_by_User}): on Medium, \emph{JavaScript} is the only programming language in the top 10 tags; on Stack Exchange, related technologies (\emph{jQuery} and \emph{Node.js}) are at the top.
These results suggest that many smart contract developers have a background in JavaScript and related technologies,
which may be explained by the similarity between JavaScript and Solidity, the most widely used high-level language for smart contract development. 
\Aron{Add more discussion?}




\ClearPage
\section{Related Work}
\label{sec:related}

\SPACE{
\Aron{We can omit this paragraph}
We discuss three areas of related work:
surveys of development practices and discussions (Section~\ref{sec:related_practices});
reviews of smart contract languages, tools, and security issues (Section~\ref{sec:related_tools});
and studies of smart contract development challenges from an educational perspective (Section~\ref{sec:related_educational}).

\Aron{Let's try to shrink related work to half a page or two thirds of a page}
}

\subsubsection{Smart Contract Development Practices}
Bartoletti et al. \cite{bartoletti2017empirical} were the first to quantitatively investigate the usage of design patterns and the major categories of smart contracts,  providing a categorized and tabulated repository of data related to smart contracts.
To this end, they examined smart contract platforms, applications, and design patterns, aggregated articles about smart contracts from \url{coindesk.com}, and identified nine common design patterns used in some combination by most of the smart contracts that they found. 
Atzei et al. \cite{atzei2017survey} presented a study of security vulnerabilities in Ethereum smart contracts, based on analysis of academic literature, Internet blogs, discussion forums about Ethereum, and practical experience in programming smart contracts.
Based on their study, they provided a taxonomy for the root causes of vulnerabilities and techniques to mitigate them.
Wohrer et al. \cite{wohrer2018design} examined design patterns for smart contracts in Ethereum, focusing on two questions: which design patterns are common in the ecosystem and how they map to Solidity coding practices. 
They employed a multivocal literature review, which considered various sources from academic papers to blogs and forums about Ethereum. Their analysis  yielded 18 distinct design patterns.
Jiang et al. \cite{jiang2018blockchain} performed a preliminary study of blockchain technology as interpreted by developers and found that blockchain related questions represent a growing minority of posts on Stack Overflow. The most common problems with blockchain are related to configuration, deployment, and discussion, followed by ten less common categories. 
However, they did not consider the development of smart contracts.



\subsubsection{Smart Contract Security Issues and Tools}    
Parizi et al. \cite{10.1007/978-3-319-94478-4_6} conducted an empirical analysis of smart contract programming languages based on usability and security from the novice developers' point of view. They considered three programming languages: Solidity, Pact, and Liquidity. 
The study concluded that although Solidity is the most useful language to a novice developer, it is also the most vulnerable to malicious attacks as novice developers often introduce security vulnerabilities, which can leave the contracts exposed to threats.
More recently, in another study, Parizi et al. \cite{parizi2018empirical} carried out an assessment of various static smart contract security testing tools for Ethereum and its programming language, Solidity. Their results showed that the SmartCheck tool is statistically more effective than the other automated security testing tools. However, their study considers only the effectiveness, usability, etc. of the tools, but not whether developers use them in practice.
%
%
Groce et al. \cite{groce2019actual} summarized the results of security assessments (both manual and automated) performed on smart contracts by a security company. The authors argued that their results pertain to more important contracts (in contrast to prior surveys) since developers were willing to pay for the assessments.
Based on the results, they categorized security issues and provided statistics on their frequency, impact, and exploitability.
Li et al. \cite{li2017survey} studied a wide range of security issues in blockchain technology. They conducted a systematic examination of  security risks to blockchain by studying  popular blockchain platforms (e.g., Ethereum, Bitcoin, Monero). 
\TODO{The smart contract flaw-finding assessment done by Groce et al. \cite{groce2019actual} does not mention any other security tool except for Slither that addresses the weaknesses. Parizi et al. \cite{parizi2018empirical}, in their assessment of current static smart contracts security testing tools; showed that the SmartCheck tool is statistically more effective than the other automated tools in the study.}

\ClearPage
\section{Discussion and Conclusion}
\label{sec:concl}
Based on the volume of smart contract related discussions on Stack Exchange (i.e., Stack Overflow and Ethereum Stack Exchange) and Medium, we found that interest in smart contracts---at least from the developers' perspective---seems to have peaked in the first few months of 2018, and has been slowly declining since then. This trend also coincides with a decline in the price of ETH. 
It will be interesting to see whether this negative trend will continue into the future, or if the decline was just a temporary disillusionment after the initial hype.

We also found that
even though most smart contract related questions on Stack Exchange receive at least one answer, extended discussions that would include many answers or comments are rare.
%
The topics of smart contract related discussion on Stack Exchange seem to be dominated by a narrow stack (e.g., Solidity, Go Ethereum, Truffle, web3.js), and we observe the prevalence of similar topics on Medium. 
For example, on both sites, alternative languages (e.g., Vyper) are rarely discussed. 

We also observed limited discussion of security-related topics on Stack Exchange, which is very concerning  since many smart contracts suffer from security vulnerabilities in practice and since many developers rely on Stack Overflow and similar sites. On Stack Exchange, less than 5\% of posts mention security or vulnerabilities; while on Medium, the ratio is around 41\%. 
On Stack Exchange, re-entrancy is the most discussed vulnerability, which seems to be in large part due to the infamous ``DAO attack.'' 
\Amin{Medium has a good number of posts about security that can alleviate such concerns.}
Similarly, Stack Exchange posts rarely mention security tools. Further, security tools are even less frequently mentioned in response to question about vulnerabilities  (e.g., in conjunction with question about re-entrancy, even though some of the tools can detect re-entrancy vulnerabilities).
Fortunately, Medium has a lot more posts  that discuss security tools. We find Oyente and Mythril to be the most popular among those tools.

Finally,  studying what other topics smart contract developers discuss, we found a significant number of posts about JavaScript and related technologies (and web technologies more generally).
This suggests that many smart contract developers have background and interest in JavaScript.

\bibliographystyle{splncs04}
\bibliography{references.bib}


\end{document}